# Channels of oxygen diffusion in single crystal rubrene revealed.

Robert J. Thompson*[ab], Thomas Bennett[c], Sarah Fearn[c], Muhammad Kamaludin[a], Christian Kloc[d], David S. McPhail[c], Oleg Mitrofanov[b], Neil J. Curson[ab]

[a] London Centre for Nanotechnology, 17-19 Gordon Street, London, WC1H 0AH, UK
[b] Dept. Electronic & Electrical Engineering, UCL, Roberts building, London, WC1E 7JE, UK
[c] Dept. Materials, Imperial College London, Royal School of Mines, Exhibition road, London, SW7 2AZ, UK
[d] School of Materials Science and Engineering, Nanyang technological University, Singapore, 639798

**ABSTRACT**

Electronic devices made from organic materials have the potential to support a more ecologically friendly and affordable future. However, the ability to fabricate devices with well-defined and reproducible electrical and optical properties is hindered by the sensitivity to the presence of chemical impurities. Oxygen in particular is an impurity that can trap electrons and modify conductive properties of some organic materials. Until now the 3-dimensional profiling of oxygen species in organic semiconductors has been elusive and the effect of oxygen remains disputed. In this study we map out high-spatial resolution 3-dimensional distributions of oxygen inclusions near the surface of single crystal rubrene, using Time of Flight Secondary Ion Mass Spectroscopy (TOF-SIMS). Channels of diffused oxygen, 'oxygen pillars', are found extending from uniform oxygen inclusion layers at the surface. These pillars extend to depths in excess of 1.8 µm and act as an entry point for oxygen to diffuse along the *ab*-plane of the crystal with at least some of the diffused oxygen molecularly binding to rubrene. Our investigation of surfaces at different stages of evolution reveals the extent of oxygen inclusion, which affects rubrene's optical and transport properties, and is consequently of importance for the reliability and longevity of devices.

**Introduction**

Organic semiconductors are of interest due to many advantageous properties: an ability to be chemically tuned allowing tailoring of material properties;[1,2] broad active optical spectra making them ideal as dyes and active layers in opto-electronic devices;[2,3] the ability to be integrated within biological systems as bioelectronics;[4] and a compatibility with flexible substrates allowing development of cost efficient manufacturing methods alongside novel device architecture.[4–6] They also offer the potential for more sustainable electronic technologies allowing biodegradable or recyclable devices.[7] However in spite of their potential impact and impressive recent developments in organic device science,[8,9] widespread commercial use is still limited by longevity and variation in device performance. When asking what limits the performance of organic devices, it is important to consider how the introduction of defects or chemical impurities affect charge transport and optical properties during a device's operational lifetime. Oxygen is a particularly important defect as it can be incorporated during both material growth, sample preparation and via interaction with the ambient environment during a device's operational lifetime. In this study we will investigate oxygen in a single crystal organic semiconductor, rubrene, which exhibits excellent optical and electronic properties, but in which oxygen incorporation is ill-understood and debated within the literature.[10–19]

In this paper we map out 3-dimensional distributions of oxygen inclusions near the surface of rubrene in order to determine the uniformity of oxide distribution and how this should inform the design of electronic devices and analysis of surface measurements. Oxide mapping is undertaken using high-spatial resolution Time of Flight Secondary Ion Mass Spectroscopy (ToF-SIMS), for the first time.

Our most important and surprising result is the observation of oxygen diffusion channels, 'oxygen pillars', extending perpendicularly from the surface into the material, to depths in excess of 1.8 μm along the c-axis of the crystal. These pillars extend from a uniform oxygen inclusion at rubrene's surface. The most startling implication of this result is that one cannot treat the material as a laterally uniform substrate for device fabrication, over the length-scale of current prototype testbed devices. Also highly significant is that we find that the pillars act as an entry point for oxygen to diffuse along the ab-plane of the crystal, with at least some of the diffused oxygen molecularly bound to rubrene as opposed to ionic or molecular oxygen sitting interstitially between molecules. The implication here is that the initial density of pillars could affect the rate at which the material degrades over time due to incorporation of oxygen from the ambient. We pursue this issue further by describing the rubrene surface at different stages of evolution, shedding light on how the oxygen inclusions vary with time.

In-order to investigate chemical impurities induced by the environment a tool is required that can directly probe the spatial chemical composition of a material. ToF-SIMS is a unique tool that offers sub-micron lateral resolution and an ability to produce 3-dimensional chemical mapping. Facilitated by the advent of large cluster ion sources the technique is now able to analyze materials consisting of larger organic molecules.[20]

The ideal materials to investigate the variation of properties due to environmental exposure are single crystal organic semiconductors. These have proved an excellent model test bed in studying the intrinsic properties of small molecule organic semiconductors. In particular, rubrene has been a source of much investigation and in single crystal form grown via physical vapour transport (PVT)[21] it provides a high purity test subject for the incorporation of defects. Single crystal rubrene displays record-high electronic and opto-electronic properties for bulk organic semiconductors with charge carrier mobility reproducibly reaching up to 20 $cm^2V^{-1}s^{-1}$ [22–24] alongside long exciton lifetimes and extraordinarily large exciton diffusion lengths.[25–27] The issue of environmental effects on rubrene is a prevalent one with oxygen incorporation and its effects poorly understood and still a source of debate within the literature. Mitrofanov et al. associated a peak in the photo luminescence spectrum ~0.25 eV below the exciton peak with an oxygen-induced electronic state.[10] They subsequently found that this peak is pronounced strongly in crystals with larger defect densities.[11] These studies agreed with Krellner et al. who observed a hole trap state 0.27 eV into the band gap after exposure to oxygen.[12] In contrast, Chen et al. associated this photoluminescence peak as due to pockets of amorphous rubrene,[13] and Song et al. estimated the HOMO of oxidised rubrene at lower energy, ~1 eV, below that of pristine rubrene.[14] In other studies Zhang et al. and Maliakal et al. concluded that the high reported mobility in rubrene is due to a raised number of holes induced by oxygen incorporation.[22,28] Najafov et al. though conversely found that the oxygen incorporation reduces both dark- and photo-conductivity of rubrene.[15] Furthermore, the mechanism of oxygen incorporation in rubrene crystals remains unclear. Previous studies treat the incorporation of oxygen as a spatially uniform process accelerated by the presence of crystallographic defects[11,16,17] and light.[15,18] Recent ToF-SIMS studies by our group, however, showed that oxygen incorporation at the surface of rubrene can form two types of oxides:[16] one preferentially incorporating at the site of structural defects while the other formed a uniform surface layer. More recently Mastrogiovani et al. reported the depth of penetration of oxygen, which was found to exceed 100 nm in 24 hours under illumination.[18]

## Experimental methods

Single crystal rubrene was grown by the vapour transport method ensuring a high purity of sample.[21] A typical crystal can be seen in previous reports from our group along with a diagrammatic representation of the crystalline packing.[16,17] This study looks at the *ab*-plane surface. Crystals were mounted on an aluminium substrate using carbon paint. Before mounting, the aluminium oxide was removed from the substrate using abrasive paper ensuring good conduction between sample and substrate, reducing the build-up of charge due to ion bombardment in ToF-SIMS analysis.

In order to study oxygen content deep within the bulk the crystal was cleaved along the *ab*-plane. The cleaving technique, detailed in previous publications from our group[17,29] produces a new pristine surface from within the bulk. In the current study the cleave produced new surfaces at a depth of 90 μm below the surface as measured using a coherent scanning interferometer.

ToF-SIMS analyses were carried out using an IONTOF SIMS V instrument equipped with a Bi Liquid Metal Ion Gun rastered to produce 2D maps and an Ar cluster ion gun used to sputter molecular layers, producing depth resolution. Surface analyses were carried out using a 25keV $Bi_3^+$ primary ion beam with a current of 0.2 pA. Depth profiles of the rubrene were obtained by sputtering the material with a 5keV $Ar^+1000$ cluster ion beam, with a beam current of 3nA, and analysing the sputtered surface using the $Bi_3^+$ primary ion beam . The sputter area to analytical area ratio was adjusted such that each $Ar^+_n$ sputter cycle removed any damage caused by the analytical primary ion beam.

Pillar diameters reported in Figure 3 were measured by the following method. Each row of pixels that intersected a specific pillar was summed together producing a single line scan in x. The line scan displayed a peak at the location of the pillar that was fitted as a Gaussian; the width in the *x*-axis was defined as the full width half maximum of this fitted Gaussian. This process was repeated in the *y*-axis. The average width of the pillar was defined as the mean of the FWHM of the Gaussian fitted in the *x*-axis and *y*-axis.

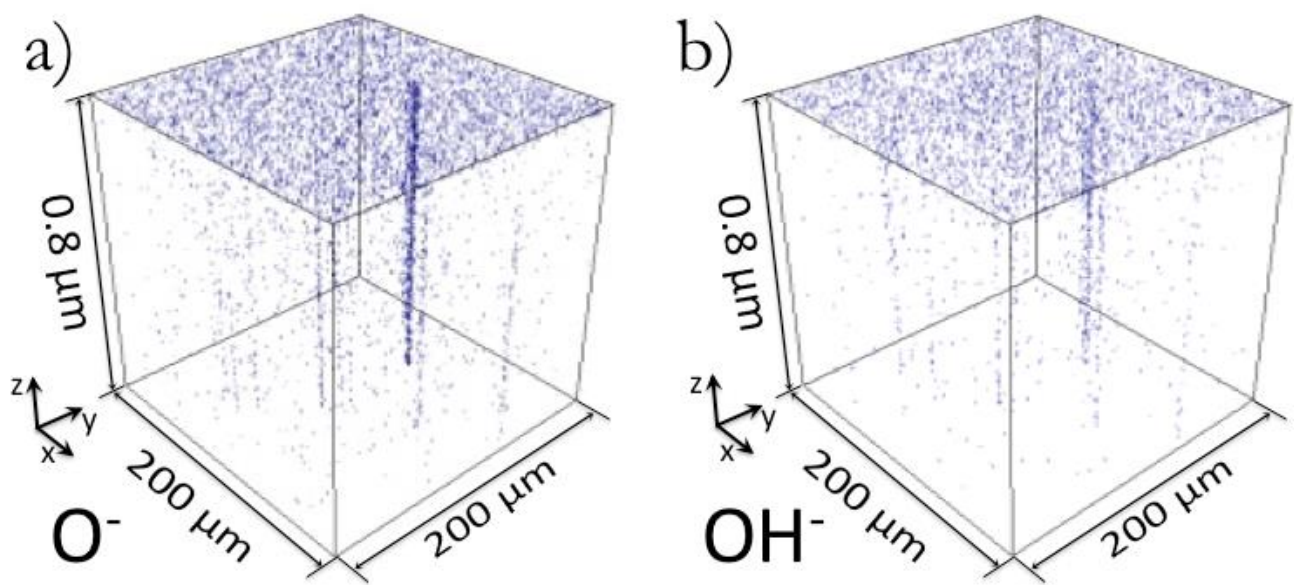

**Figure 1.** 3D rendered plots showing the spatial distribution of $O^-$ and $OH^-$ ions in figures a) and b) respectively. Transparent blank pixels indicate none of the respective ionic species were present at that location. A uniform surface coverage of $O^-$ and $OH^-$ can be seen along with oxygen "pillars" showing a preferential diffusion of oxygen along these channels in the c-axis. The x,y plots of this data integrated over the z axis can be seen in Figures 2 a) and b).

**Results and Discussion**

3-dimensional plots of the spatial distribution of $O^-$ ions and $OH^-$ ions within a 3-month old sample are shown in **Figure 1**. This surface was not specially prepared post growth and is termed 'as grown'. The ToF-SIMS provided mass spectral depth profiling, in this case up to a depth of 0.8 µm. The plots in Figure 1 track the peaks in the mass spectra that represent the mass per charge (m/z) of the ions $O^-$ and $OH^-$. Figure 1 clearly shows there is a greater density of $O^-$ and $OH^-$ ions at the surface creating a uniform surface layer. At the interface between this layer and the bulk of the material, the distribution of oxygen becomes concentrated in channels (pillars) running in the *c*-direction of the crystal perpendicular to the surface. We observed such pillars in all (five) 'as grown' samples investigated to date, with the minimum density observed to date being 2 pillars in a 200 x 200 µm² analysis area (50 cm⁻²) and extend to depths of at least 1.8 µm. The pillars whose origin is discussed below are not due to any artefact caused by a slower sputter rate as determined by the lack of topographical features seen in AFM images before and after ion beam analysis.

In order to determine the nature of the pillars **Figure 2** compares the distribution of several molecular species. Figures 2 a) to f) present the distribution of secondary ions compressed to *xy*-plots with each column of pixels in the z direction integrated into a single pixel. Figures 2 a) and b) show the distribution of $O^-$ and $OH^-$ as taken from Figure 1, while c) to f) track the distribution of $C_{42}H_{28}^-$, $C_{42}H_{28}O^-$, $C_{42}H_{28}O_2^-$ and $C_2H^-$ respectively. $C_2H^-$ is a molecule not expected to be found intrinsically within rubrene, consequently its high count is indicative of fragmentation occurring when large 'parent' molecules are liberated from the surface. $C_{42}H_{28}^-$ is the intact rubrene molecule and would be expected to show a uniform covering in an intrinsic material containing no defects. $C_{42}H_{28}O^-$ and $C_{42}H_{28}O_2^-$ correspond to rubrene oxide and rubrene peroxide respectively. Previous studies have shown that these oxide species of rubrene bind to different sites on the surface with rubrene peroxide preferentially being found at the sites of defects and rubrene oxide forming a more uniform covering of the surface.[16] As can be seen in Figure 2 there is a clear correlation between each of the ionic species tracked and the locations of the pillars. Several of the pillars can be identified even in the low count rubrene peroxide map. Arrows highlighting three of the largest pillars are shown in each image. In general there is a reduction in count for the larger molecules at the location of the pillars and an increase in count of the smaller mass species. The increased smaller molecular count implies that at the location of the pillars there is a raised probability of the molecule fragmenting when liberated from the surface. A change in the fragmentation probability such as this implies a variation in the relative bond strengths. This variation can be introduced by a structural defect causing a variation in intramolecular bonding, or a molecular variation, such as oxidation, that will cause both inter- and intra-molecular variation. A physical defect such as a threading dislocation will induce structural deformation of the lattice allowing molecular conformations that are known to facilitate oxidation of rubrene. We therefore believe the appearance of the pillars is due to oxygen diffusion into the material via structural defects within the rubrene. Observations of pillars extending in the *c*-axis of the crystal are consistent with oxygen incorporation facilitated by a threading dislocation. Such dislocations are often introduced within the crystal lattice during growth in order to relieve strain.[30]

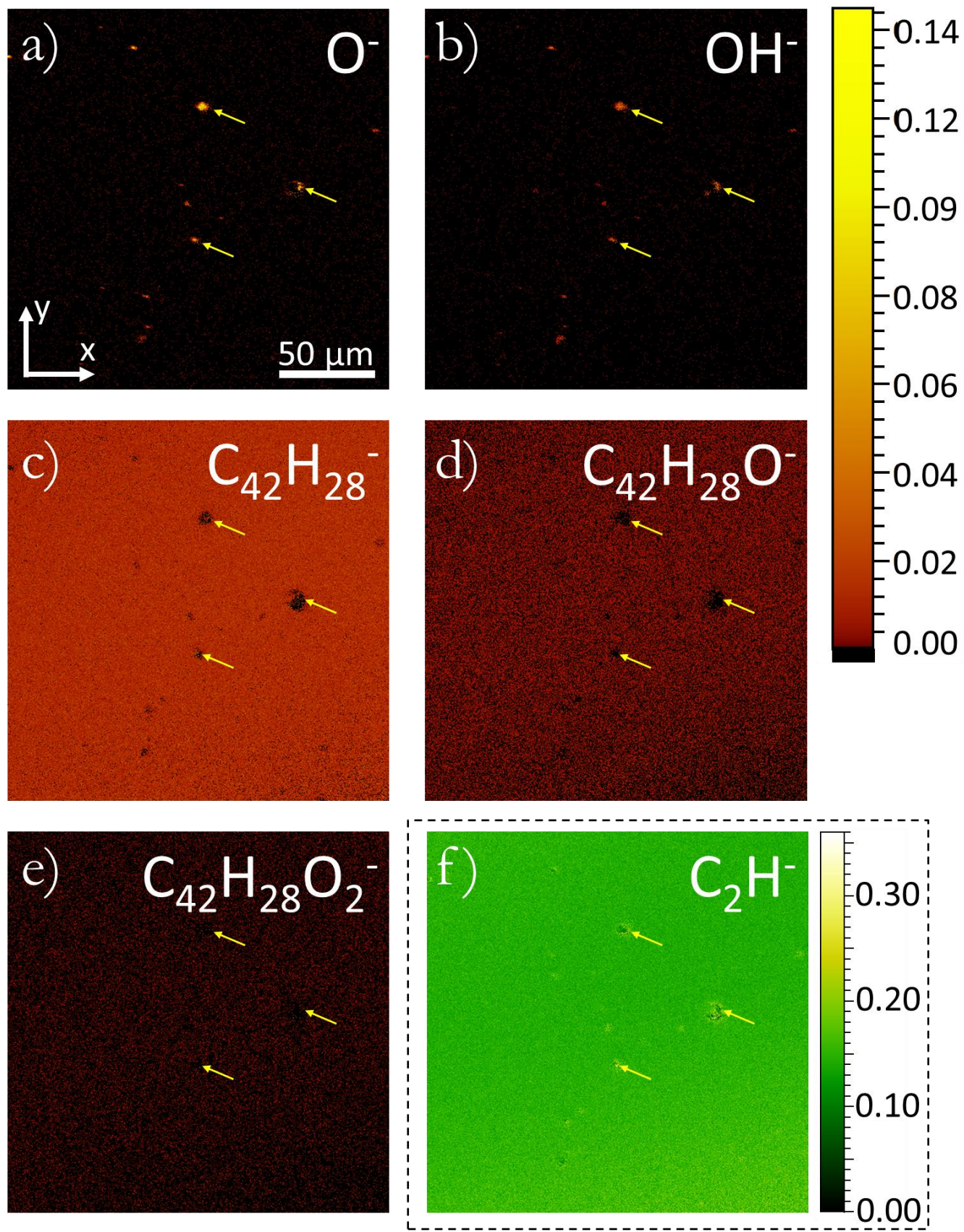

*Figure 2.* 3D x,y plots of the distribution of secondary ions normalised by the total ion count of each respective pixel. Ions O⁻, OH⁻, $C_{42}H_{28}^-$, $C_{42}H_{28}O^-$, and $C_{42}H_{28}O_2^-$, $C_2H^-$ have been plotted in Figures a), b), c), d), e) and f) respectively. All plots use the same colour scale except f) which due to a larger ion count uses the adjacent colour scale. In each plot data has been acquired in three dimensions with

*every column of pixels in z being integrated to create a 2D plot. The xy plane of these plots lies in the ab-plane of the crystal. Figures a) and b) are plots of the same data shown in figures 1 a) and b), orientation legends are included for ease of comparison. Arrows are repeated in each image showing the location of three of the largest pillars.*

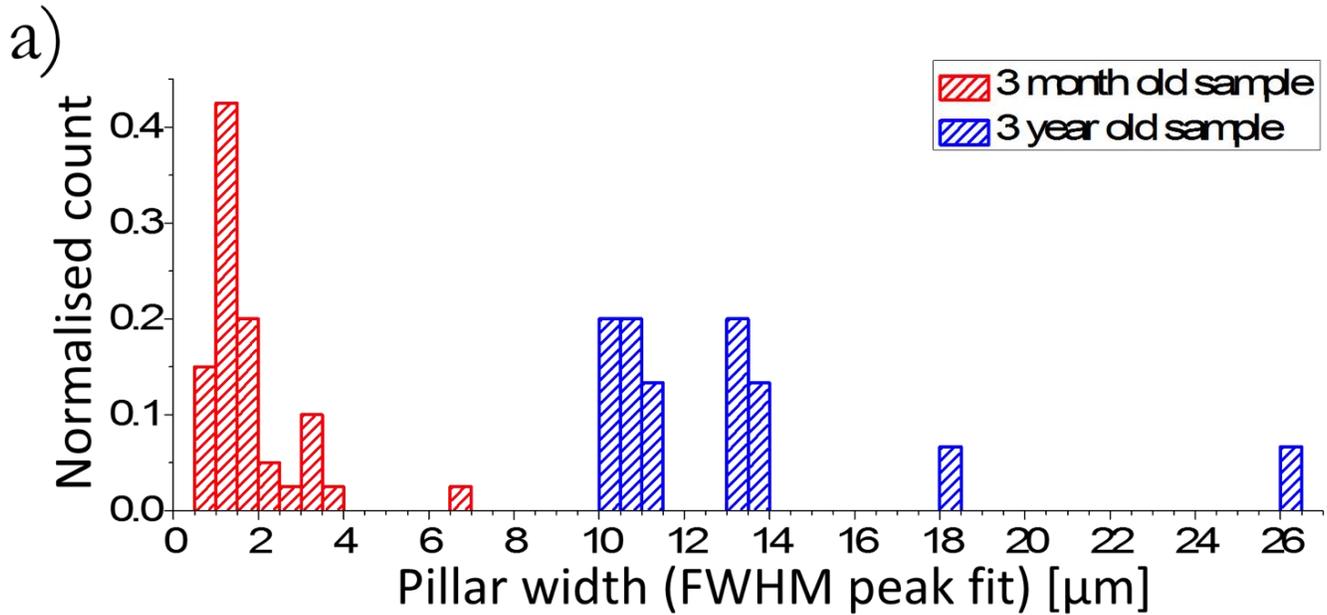
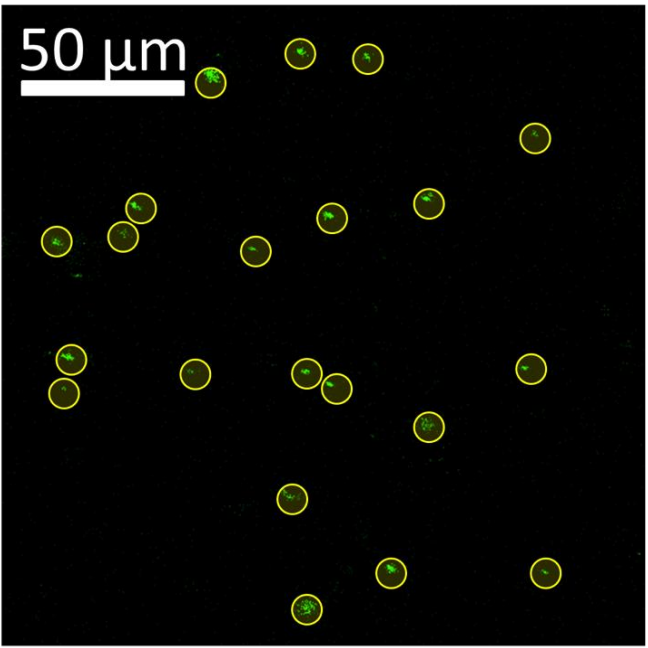
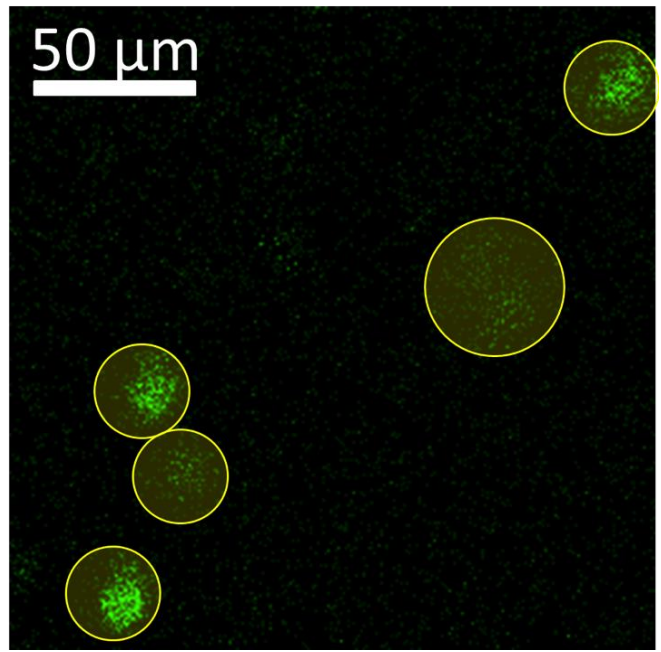

**Figure 3.** a) Histogram showing the distribution of pillar widths in a 3-month unintentionally oxidized sample and a 3-year old sample exposed to light and ambient conditions. (b) and (c) show typical xy ion maps tracking the O- ions for the 3 month and 3-year-old samples respectively. All pillars have been highlighted. The method of measuring pillar widths is described within the experimental details.

In order to test the hypothesis that defects act as the entry point for oxygen inclusion, a sample that had been exposed to ambient laboratory conditions for a prolonged period was analysed. This sample was grown and prepared in the same conditions as the crystal presented in Figure 1 and 2 with the exception that this 3-year-old sample had been intentionally exposed to light in ambient laboratory conditions over the course of 12 months before analysis. In this sample the pillar density, which vary between samples, was lower than that of the 3-month-old sample. Pillars in the aged sample were seen to have an increased diameter. **Figure 3** shows a comparison of the two samples with Figure 3 a) showing a histogram of pillar width in the two samples. The data set used within the histogram was acquired from two 200 x 200 µm$^2$ areas of the 3-month-old crystal providing 40 analysed pillars, and three 200 x 200 µm$^2$ areas of the aged sample providing 16 analysed pillars. Each data set was normalised by the total count for the respective sample. Figure 3 b) and c) show the *xy*, *z*-integrated, plot tracking O$^-$ ions for a 200 x 200 µm$^2$ area of each crystal, in each case all the pillars identified are highlighted. The increased diameter of the oxygen pillars seen in the 3-year-old sample implies that the structural defects do not only allow oxygen to penetrate along the defect but also acts as an entry point for the oxygen to diffuse from the defect along the *ab*-plane of the crystal.

Having analysed the incorporation of oxygen at the surface and sub-surface, knowledge of oxygen incorporation in the deep bulk was sought. The three-month-old sample was cleaved along the *ab*-plane in ambient laboratory conditions (see experimental section) to produce a new *ab*-plane surface 90 µm below the original surface. The sample was then immediately placed in the vacuum load-lock of the ToF-SIMS instrument. Less than 20 seconds elapsed between cleaving and evacuating. The analysed cleaved surface, at a depth of 90 µm, showed no observable indication of pillars, instead exhibiting a homogeneous chemical distribution with a much reduced oxygen count. The lack of observable pillars immediately after cleaving along with low oxygen content shows the high quality of the crystals with an extremely low impurity content. It also allows us to conclude that the depth of oxygen diffusion into this sample was less than 90 µm.

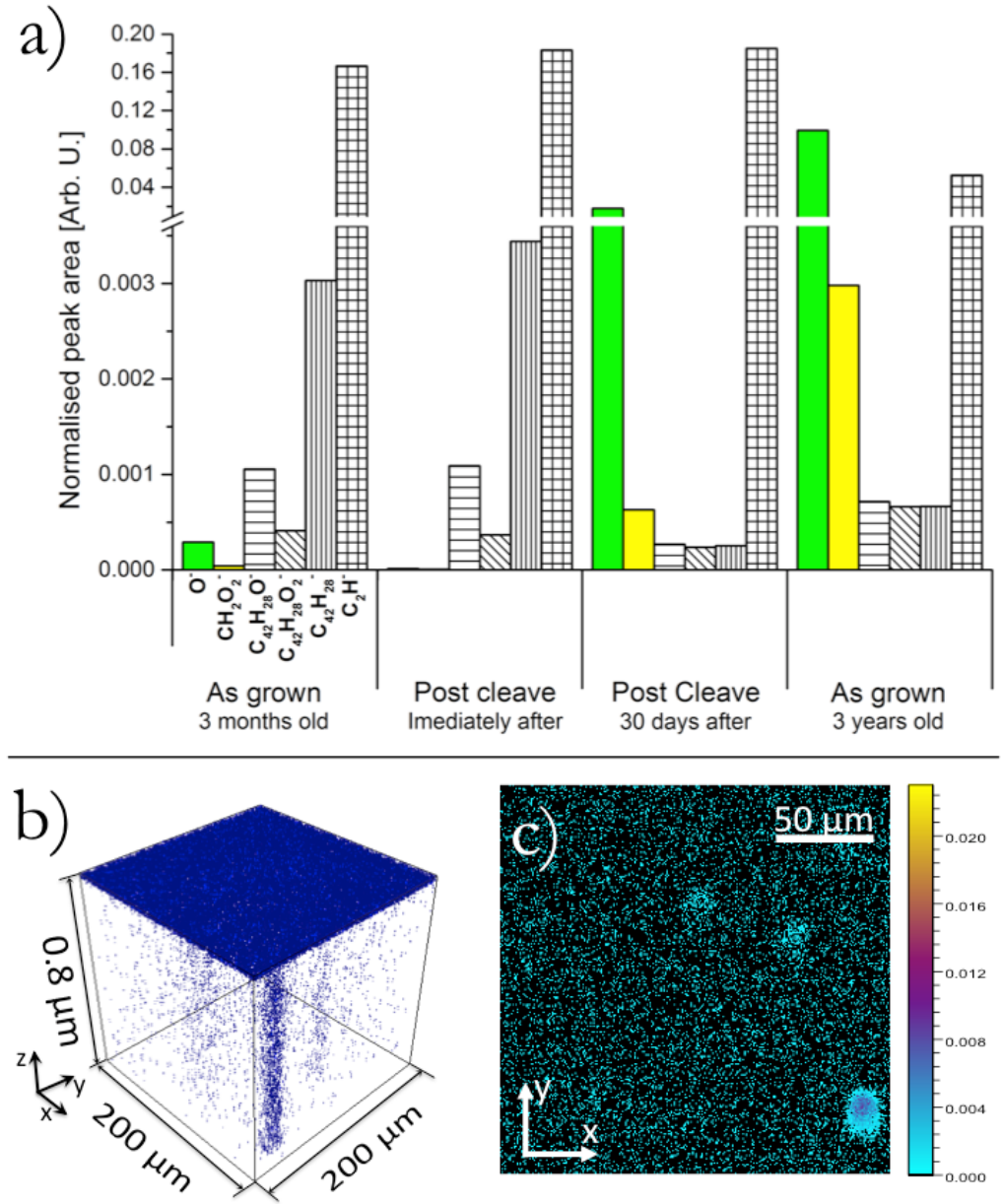

***Figure 4.*** *a) Bar chart showing the abundance of labelled secondary ions liberated from each of the labelled sample surfaces. For clarity the y-axis is broken with a change in scale allowing comparison of both low and high abundance molecular components. It should be noted that no Ar cluster sputtering was used in these surface measurements and therefore the relative abundance of molecular components is only representative of the top molecular layers. Figures b) and c) show the local variation of $CH_2O_2^-$ abundance in the 3 year old 'as grown' sample. Ar cluster ions were used to allow depth profiling. Figure b) is a 3D rendered plot and c) a xy plot with integration in Z.*

While we have determined that there are no oxygen pillars in the cleaved rubrene sample, 90 µm below the surface, the question of whether the structural defects with which they are associated are also absent was investigated by leaving the sample in ambient laboratory conditions for 30 days. The cleaved surface was then re-analysed at a location not previously exposed to the ion beam. Although the sample showed an increased oxygen count at the cleaved surface in the form of a uniform layer, there were still no observable signs of pillar development. With oxygen having diffused into the surface it would be expected that any dislocations existing in the crystal at the depth of the cleaved surface would begin to be decorated by oxygen, allowing them to be observed. The lack of any observable pillars implies that dislocations did not exist at this depth within the crystal. This could imply that the pillars are a result of strain experienced later in the growth of the crystals that is not present in earlier stages when deeper parts of the crystal are deposited.

To study the concentration of different species at each of the four surfaces previously discussed, each surface was analysed by ToF-SIMS without the use of the heavy ion sputter beam. By ensuring that in this situation the analysis beam did not exceed the static limit (within the static limit the number of primary ions is low enough that statistically only the first monolayer is removed), mass spectra were only acquired from the top molecular layers. **Figure 4** a) displays this surface sensitive mass spectra, where the bar chart shows the normalised secondary ion count of the labelled ions liberated from each respective surface. In each case the count is normalised to the total secondary ion count of the respective analysis and is averaged over a 200 µm$^2$ area.

Looking firstly at the O$^-$ content in each surface it can be seen that the three month old 'as grown' surface has a relatively low content which is reduced even further by a factor of 10 when the sample is cleaved revealing the bulk molecular makeup. It is indicative of the high quality of the sample that a low level of oxygen was incorporated at the surface and within the bulk there is only a trace presence of oxygen. Upon exposing this cleaved surface to the environment the oxygen content increased dramatically by three orders of magnitude from an average normalised count of 1.2×10$^{-5}$ immediately after cleaving to 1.8×10$^{-2}$ 30 days later. This agrees with previous work from our group that showed that cleaving of the sample produced a more reactive surface.[17] The aged 'as grown' surface shows an even greater oxygen count of 0.10, this sample had been exposed to ambient condition and light as expected this sample shows the highest level of oxygen content which is in agreement with work by Najafov et al. and Mastrogiovanni et al.[15,18] The $CH_2O_2^-$ peak follows the same trend as the oxygen ion. As with the $C_2H^-$ ion this molecule is not expected to exist within the sample and is therefore attributed to fragmentation of a larger 'parent molecule' the most likely candidate being rubrene peroxide. Due to the low $CH_2O_2^-$ count in the three-month-old sample the molecule was not initially identified as obviously having any spatial correlation with pillars. Figure 4 b) and c) plot the location of $CH_2O_2^-$ ions within the three year old aged sample, b) is a 3D rendered plot and c) an xy, z-integrated plot. Figure 4 b) and c) clearly show that this molecular species is localised at the position of pillars. The increase in $CH_2O_2^-$ alongside that of atomic oxygen shows that the pillars incorporate oxygen bonded to the rubrene molecule rather than just atomic or molecular oxygen sitting within the defect.

## Conclusion

In this work we have uncovered preferential channels of oxygen diffusion into single crystal rubrene, in the form of 'pillars' of oxygen extending from the surface of the crystal along the *c*-axis. These pillars are likely to exist because of the presence of structural defects formed during crystal growth. ToF-SIMS spectra imply that some or all of the constituents of the pillars are an oxide of rubrene as opposed to ionic or molecular oxygen. Analysis of crystals exposed to ambient conditions shows that as well as allowing channels for oxygen to penetrate along the *c*-axis of the crystal, these pillars act as an entry point for oxygen to diffuse along the *ab*-plane of the crystal. This diffusion along the *ab*-plane has the effect of increasing the diameter of these oxygen pillars beyond the initial point like structural defect. By cleaving the crystal to expose new surfaces 90 µm below the 'as grown' surface it was found that structural defects creating the oxygen channels do not exist deep into the crystal and therefore are likely to be incorporated during the later stages of growth. Finally we provide a comparison of the surface inclusion of oxygen for four different rubrene surfaces with different histories. The properties of these surfaces vary considerably, implying that a systematic sample preparation process is required for attaining reproducible surface properties. One appealing solution is sample cleaving, as it produces clean, flat surfaces with wide terraces and low oxygen content in the region below the surface. However, the fact that cleaved surfaces have enhanced reactivity would need to be understood. In electronic devices where the active components are at the surface of the material, such surface preparation protocols will be a prerequisite for the fabrication of reliable devices.

## Acknowledgments

Dr. R Thompson is funded by the EPSRC doctoral prize fellowship.